\documentstyle[preprint,epsbox]{jpsj}

\normalsize

\newcommand{\beqa}{\begin{eqnarray}}
\newcommand{\enqa}{\end{eqnarray}}
\newcommand{\beq}{\begin{eqnarray}}
\newcommand{\enq}{\end{eqnarray}}
\title
{Three-leg Antiferromagnetic Heisenberg Ladder \\
with Frustrated Boundary Condition; \\
Ground State Properties
}

\author
{Kenro {\sc Kawano} and Minoru {\sc Takahashi}}

\inst
{Institute for Solid State Physics, \\
University of Tokyo,Roppongi,Minato-ku,Tokyo 106 Japan}

\recdate
{
\today
}

\abst
{The antiferromagnetic Heisenberg spin systems on the 
three-leg ladder are investigated. Periodic boundary 
condition is imposed in the rung direction. The system has an excitation gap for all
antiferromagnetic inter-chain coupling ($J_{\perp}>0$). The estimated gap for
the strong coupling limit ($J_{\perp}/J_1 \rightarrow \infty$) is 0.28$J_1$. Although the
interaction is homogeneous and only nearest-neighbor, the ground states
of the system are dimerized and break the translational
symmetry in the thermodynamic limit.
Introducing the next-nearest neighbor coupling ($J_2$), we can see
that the 
system is solved exactly. The ground state wave function is
completely
dimer-ordered.
Using density matrix renomalization group algorithm, we show numerically
that the original model ($J_2=0$) has the same nature with the exactly
solvable model.  
The ground state properties of the ladder with a higher odd number of legs are also discussed.
}

\kword
{odd-legs ladder,dimerization,frustration}

\begin{document}
\sloppy
\maketitle

\pagebreak
\section{Introduction}
Nowadays there is much attention for quantum ladder systems.\cite{Dagotto}
One of the reason is that the ladder systems are the first step from
one-dimensional toward two-dimensional.\cite{Chakraverty,Schulz2,Lin}
The other reason is caused by the advance of
experimental techniques.
Many materials which seem to be ladders  are found and investigated
extensively.\cite{Hiroi,Azuma,Kojima}
The antiferromagnetic spin systems are  among the simplest 
many body systems which
include
strong quantum effect.
From the analysis using the $O(3)$ non-linear
$\sigma$ model,\cite{Sierra}
antiferromagnetic spin ladders with an even
number of legs are expected to have an excitation 
gap. In this case, as for the existence of an excitation gap,
the system is not sensitive to the boundary condition perpendicular to 
the chain. 
For the two-leg $S=1/2$ spin ladder there are many works and it is 
confirmed that there is an excitation
gap.\cite{Strong,Watanabe,White30} On the other hand, in the case of a 
ladder with an odd
number of legs, the topology 
is important.\cite{Schulz} (For the fermion systems, there are works
about the importance of the topology,\cite{Lin,Arrigoni}
which treat the repulsive Hubbard ladder with weak electron-electron coupling limit using field theoretical argument.)
The odd-leg ladders with open boundary condition in the rung direction  (we will call it as open ladder) has no frustrated interaction. These can be considered 
effectively as $S=1/2$ antiferromagnetic Heisenberg chain (AFHC)\cite{Sierra,Frischmuth}. On the other hand, periodic boundary condition in the
rung direction (cylindrical ladder) causes
frustrated interaction. The frustration changes the situation dramatically.
There is also attention for the dimer-ordered  phase.  We know the
$S=1/2$ AFHC is unstable for the dimer
interaction.  The model becomes gapful as the dimer interaction is inserted.\cite{AGSZ}
However the model breaks the translational invariance a priori. The Majumdar-Ghosh
model is one of the typical models with which ground states  are  frustration induced dimer-ordered 
and have translational invariance in the finite size system.\cite{MG} In this paper we show
another example of the model with frustration induced dimer-ordered
ground states.   
We consider following antiferromagnetic 
Heisenberg Hamiltonian,
\beq
{\cal H}_{1}=J_{1}\sum_{n=1}^{N'}\sum_{i} H_{n,n}^{i,i+1}
         +J_{\perp}\sum_{n=1}^{N'}\sum_{i} H_{n,n+1}^{i,i},\label{Hamil1}
\enq
where $ H_{n,m}^{i,j}=\mbox{\boldmath $S$}_{n,i} \cdot \mbox{\boldmath
$S$}_{m,j}$ and $\mbox{\boldmath $S$}_{n,i}$ is $S=1/2$
 $SU(2)$ spin operator.
We concentrate on the condition, $J_1,J_{\perp}\geq 0$.
Subscripts and superscripts $n,m$ and $i,j$ represent the site number in the rung direction  and the chain direction respectively. The number of 
legs and the number of sites along the chain are $N$ and $L$
respectively. (The total number of sites is $NL$.) 
For the open ladder we set $N'=N-1$ and for the cylindlical ladder
we set $N'=N$ and $N+1\equiv 1$.
The boundary condition along the chain direction is appropriately treated 
and denoted periodic boundary condition and open one as PBC and OBC respectively.
In almost all part of this paper (except for \S  4), we will consider
the three-leg ladder
$N=3$. We add the next-nearest
neighbor interaction,
\beq
{\cal H}_{2}=J_{2}\sum_{n=1}^{N'}\sum_{i} H_{n,n}^{i,i+2},\label{J2}
\enq
to see the system more explicitly. (See below.)
Then the total Hamiltonian is ${\cal H}={\cal H}_{1}+{\cal H}_{2}$.
In the case of $J_{\perp}=0$, we know the model well for a few parameter points. 
The model with $(J_1,J_{\perp},J_2)=(J_1,0,0)$, which
is exactly solved by Bethe Ansatz,\cite{Hulthen} 
has no frustrated interaction and ground state is proved to
be spin singlet by Marshall-Lieb-Mattis Theorem. 
The model  has gapless excitation and belongs to the universality class
of the level-1 $SU(2)$ WZW model in the
low energy limit. 
We know the long-distance asymptotic behavior of 
correlation functions, for example, spin-spin correlation function 
$\langle\mbox{\boldmath $S$}_0\cdot \mbox{\boldmath $S$}_r\rangle$
 and dimer-dimer correlation function 
$\langle(\mbox{\boldmath $S$}_0\cdot \mbox{\boldmath $S$}_1)(\mbox{\boldmath
$S$}_r
\cdot \mbox{\boldmath $S$}_{r+1})\rangle-\langle\mbox{\boldmath $S$}_0\cdot \mbox{\boldmath $S$}_1\rangle\langle\mbox{\boldmath
$S$}_r
\cdot \mbox{\boldmath $S$}_{r+1}\rangle$  decay as $(-1)^r /r$ 
(up to logarithmic corrections\cite{GSchulz}).
For the other parameter point $(J_1,0,J_1/2)$, which is the so-called Majumder-Ghosh
model,\cite{MG} the ground states are  well-known in the sense that we exactly know the 
wave function. The ground state wave function is dimerized and  written down by
direct products of two-site spin-singlet.  Therefore the ground states
are doubly degenerate and break the translational invariance. 
The existence of the excitation gap is exactly proved.\cite{AKLT}
By the bosonization approach, we know that the perturbation by  $J_2$ is
marginal. (Whether the operator is relevant or irrelevant is determined 
by the sign of the initial coupling constant. In the decoupled  chains case 
($J_{\perp}=0$), the transition point is
$J_2=0.2411\cdots$.\cite{Nomura}) On the other hand, perturbation by
$J_{\perp}$  
is strongly relevant
compared with that of $J_2$.\cite{Schulz,Totsuka} Therefore we expect
that the total Hamiltonian 
with $J_{\perp} >0$ is essentially represented by the effective Hamiltonian with
$J_{\perp}/J_1 \rightarrow \infty$.

Our strategy to investigate the model is as follows.
First, we introduce the effective Hamiltonian in the strong coupling
limit $(J_{\perp}\rightarrow \infty)$, which is the fixed point of 
the renomalizaion group.
We add the next-nearest neighbor coupling $(J_2)$ to the model.
Then we show the model with special coupling $(J_2/J_1=1/2,J_{\perp}/J_1
\rightarrow \infty)$ is solved exactly. (In the sense that we can construct
the exact ground state wave function.)
Using the density matrix renomalization group (DMRG) algorithm introduced
by White,\cite{White} we see the original model $(J_2=0$ and $ 0 < J_{\perp}/J_1 \leq\infty)$
has the same nature with the exactly solved model by calculating the
expectation value of the local Hamiltonian operator. We also confirm that there is no transition between $0\leq J_{2}/J_1\leq 1/2$ through seeing no level
crossing in low-energy excitation spectra by numerical exact diagonalization.  
The estimation of the excitation gap for various value of $J_{\perp}$
is also showed.
In $\S 4$  we shortly mention the cylindrical ladders with a higher odd number of legs.
For general odd number of legs $N$, we construct effective Hamiltonian in the
strong coupling limit, which has the same form with the case of $N=3$.
\section{Strong-Coupling Effective Hamiltonian}
We first consider the strong coupling limit ($J_{\perp}\rightarrow
\infty$) of the three-leg ladder.  The reason is as following.  In bosonization language (weak 
coupling analysis), the interchain coupling causes the relevant
operator 
through the staggered part of the spin operator,\cite{Schulz,Totsuka} so
generally the system is expected to flow into the strong
coupling limit in renormalizaion processes, i.e. the strong coupling
limit is the fixed point of the renomalization flow. 
In the case of the open ladder with $J_1=0$ (which is understood as $
J_{\perp}=\infty$), the ground state of each 
rung is a doublet. Then the effective Hamiltonian of the original
Hamiltonian (\ref{Hamil1}) in 
the strong coupling limit is represented by,
\beqa
{\cal H}_{{\rm eff}}=J_{1} \sum_{i}\mbox{\boldmath $S$}_{i} \cdot
\mbox{\boldmath $S$}_{i+1}.\label{usualeff}
\enqa
Therefore, in this limit, the system
can be described by $S=1/2$ AFHC as noted in the introduction. The 
model has no excitation gap, power-law decaying spin-spin correlation, 
and it is classified into the same universality of level-1 $SU(2)$ WZW model.\cite{AH}
The vanishing excitation gap is proved by Affleck and
Lieb.\cite{Lieb} 
On the other hand, in the case of the cylindrical ladder, we cannot
prove the existence of a gapless excitation by the theorem
%the theorem cannot be applicable 
because the uniqueness of
the ground state is not proved.
The ground state of the three-site ring is 
four-fold degenerate, these states are,
\beqa
|\uparrow L \rangle&=&\frac{1}{\sqrt{3}}\left(|\uparrow \uparrow
\downarrow\rangle +\omega |\uparrow \downarrow \uparrow \rangle+\omega^{-1} |\downarrow 
\uparrow \uparrow\rangle\right), \nonumber \\ 
|\downarrow L \rangle&=&\frac{1}{\sqrt{3}}\left(|\downarrow \downarrow
\uparrow\rangle +\omega |\downarrow \uparrow \downarrow \rangle+\omega^{-1} |\uparrow 
\downarrow \downarrow\rangle\right),\nonumber \\ 
|\uparrow R \rangle&=&\frac{1}{\sqrt{3}}\left(|\uparrow \uparrow
\downarrow\rangle +\omega^{-1} |\uparrow \downarrow \uparrow \rangle+\omega |\downarrow 
\uparrow \uparrow\rangle\right), \nonumber \\ 
|\downarrow R \rangle&=&\frac{1}{\sqrt{3}}\left(|\downarrow \downarrow
\uparrow\rangle +\omega^{-1} |\downarrow \uparrow \downarrow \rangle+\omega |\uparrow 
\downarrow \downarrow\rangle\right),\nonumber 
\enqa
where $\omega = \exp (\frac{2 \pi {\rm i}}{3})$. The indices $L,R$ 
represent the momentum of the three site ring, $k=2\pi/3$,
and $-2\pi/3$ respectively. The four-fold degeneracy  essentially
distinguishes the cylindrical ladder from the open ladder. 
The other states have higher energy of order $J_{\perp}$.
Therefore in the strong coupling limit $J_{\perp}/J_1 \rightarrow \infty$, the effective Hamiltonian is,\cite{Schulz}
\beqa
{\cal H}_{{\rm eff}}=\frac{J_{1}}{3} \sum_{i}H_{{\rm eff}}^{i,i+1}(\alpha) \quad {\rm with} \quad \alpha=4.\label{hameff}
\enqa
where $H_{{\rm eff}}^{i,i+1}(\alpha)=\mbox{\boldmath $S$}_{i} \cdot
\mbox{\boldmath $S$}_{i+1} \left( 1+\alpha (\tau ^{+}_{i}\tau
^{-}_{i+1}+ \tau ^{-}_{i}\tau
^{+}_{i+1})\right)$. 
Though we write $\alpha$ explicitly for convenience through this
section, it should  be considered as 4.
The operators $\tau^{\pm}$  exchange indices $L$,$R$ such as,
\beqa
\tau^{+}|\cdot L\rangle&=&0, \quad \quad \quad  \tau^{-}|\cdot
L\rangle=|\cdot R\rangle,
\nonumber \\
\tau^{+}|\cdot R\rangle&=&|\cdot L\rangle,\quad \tau^{-}|\cdot R\rangle=0.\nonumber
\enqa
(In these equations the dots represent $\downarrow$ or $\uparrow$.)
We define the diagonal operator $\tau^{z}$ which, with
$\tau^{x}=(\tau^{+}+\tau^{-})/2$ and $\tau^{y}=(\tau^{+}-\tau^{-})/2i$, satisfies commutation
relation of three generators of $SU(2)$.

For the two-site system, the ground state wave function
of the Hamiltonian (\ref{hameff}) is,
\beqa
|\uparrow L,\downarrow R\rangle-|\downarrow L,\uparrow R\rangle+|\uparrow
R,\downarrow L\rangle-|\downarrow R,\uparrow L\rangle \nonumber \\
\equiv (|\uparrow \downarrow\rangle-|\downarrow \uparrow\rangle)\cdot(|LR\rangle+|RL\rangle),\label{2site}
\enqa
the corresponding energy is 
$-\frac{1}{4}(1+\alpha) J_1$. In the following, we write this wave function as,
$\bullet\!\!\!\!\rightarrow\!\!\!\!\!-\!\bullet$.
The wave function (\ref{2site}) can be  also written down in the three-leg 
ladder language.  
The wave function in the three-leg ladder is the sum of three terms which are
$[1_1,2_1]\otimes [1_2,1_3]\otimes [2_2,2_3]$ and ones translated in
the rung direction i.e.,
\begin{figure}
  \epsfile{file=threefig.epsi}
\end{figure}
\linebreak
Here $[\cdot,\cdot]$
 represents the two-site spin-singlet state $(|\uparrow
\downarrow\rangle-|\downarrow \uparrow\rangle)$. Of course the
expression is not unique, we are able to write it in more complicated form with two
site spin singlet states.

Next  we add the next-nearest neighbor coupling (Hamiltonian ${\cal
H}_{2}$). Then the total
effective Hamiltonian is,
\beqa
{\cal H}_{{\rm eff}}=\frac{J_{1}}{3} \sum_{i}H_{{\rm eff}}^{i,i+1}(\alpha)+\frac{J_{2}}{3} \sum_{i}H_{{\rm eff}}^{i,i+2}(\alpha).\label{effMG}
\enqa 
With special coupling $J_{1}=2 J_{2}$, we
find exact ground states as follows.
Define the operator $P$,
\beqa
P_{i,i+1,i+2}=\frac{1}{2}(H_{{\rm eff}}^{i,i+1}(\alpha)+H_{{\rm eff}}^{i+1,i+2}(\alpha)+H_{{\rm eff}}^{i,i+2}(\alpha))+\frac{3}{8}(1+\alpha).\label{project}
\enqa
Using this operator $P$ the Hamiltonian (\ref{effMG}) is represented by 
\beqa
{\cal H}_{{\rm eff}}=\frac{J_1}{3}\sum_{i}(P_{i,i+1,i+2}-\frac{3}{8}(1+\alpha)).
\enqa

The operator $P_{i,i+1,i+2}$ is small dimensional, we, therefore, easily proved the
semi-positive definiteness of it. The number of eigenvectors with the
eigenvalue zero is 12, and these eigenvectors are represented in
the following three types,
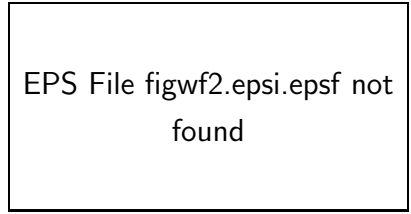
\vspace{5mm}
\begin{figure}
\epsfile{width=8cm,file=figwf2.epsi}
\end{figure}

i.e. $P$ is the linear combination of projection operators with
positive coefficients, which projects out
these three states. The situation is closely similar to the case of
the Majumdar-Ghosh model. The difference is that $P$ is not represented by 
a single projection operator.
Then the two states,
\beqa
|\psi_{1}\rangle&=&\!\!\quad\bullet\!\!\!-\!\!\!-\!\!\!\bullet\quad\!\!\!
\bullet\!\!\!-\!\!\!-\!\!\!\bullet\quad\!\!\!
\bullet\!\!\!-\!\!\!-\!\!\!\bullet\quad\!\!\!
\bullet\!\!\!-\!\!\!-\!\!\!\bullet\quad\!\!\!
\bullet\!\!\!-\!\!\!-\!\!\!\bullet\quad\!\!\!
\bullet\!\!\!-\!\!\!-\!\!\!\bullet\quad\!\!\!
\bullet\!\!\!-\!\!\!-\!\!\!\bullet\quad\!\!\! \nonumber \\
|\psi_{2}\rangle&=&\quad\!\!\bullet\quad\!\!\!\bullet\!\!\!-\!\!\!-\!\!\!\bullet\quad\!\!\!
\bullet\!\!\!-\!\!\!-\!\!\!\bullet\quad \!\!\!
\bullet\!\!\!-\!\!\!-\!\!\!\bullet\quad\!\!\!
\bullet\!\!\!-\!\!\!-\!\!\!\bullet\quad\!\!\!
\bullet\!\!\!-\!\!\!-\!\!\!\bullet\quad\!\!\!
\bullet\!\!\!-\!\!\!-\!\!\!\bullet\quad\!\!\!
\bullet\!\!\nonumber \\
&&\quad \!\!
1\quad\quad\quad\quad\quad\quad\quad\quad\quad\quad\quad\quad\quad\quad\quad\quad\!\!
L \nonumber
\enqa
are, at least, among the ground states with energy
$-\frac{1}{8}(1+\alpha) L J_1$. ( For PBC case, the site 1 and $L$ of the
wave function $|\psi_{2}\rangle$ are 
connected by a line. For OBC case, $|\psi_{2}\rangle$ is degenerate because
of the degrees of freedom of site 1 and $L$.) These ground states are evidently dimerized. In finite-size system with PBC we construct two ground states with the momentum 0 and $\pi$ as $(|\psi_{1}\rangle+ |\psi_{2}\rangle)/\sqrt{2}$ and $(|\psi_{1}\rangle- |\psi_{2}\rangle)/\sqrt{2}$. We also confirmed that only these two 
states are ground states by numerical exact diagonarization up to 12-site system.
For the two ground states, obviously, correlation functions $\langle
S^{z}_{0}S^{z}_{r}\rangle $ and
$\langle\tau^{z}_{0}\tau^{z}_{r}\rangle$ have zero correlation except
for the nearest neighbor.
The wave 
function is site local, it strongly suggests the existence of the
excitation gap.
We, unfortunately, have not found the exact proof of it.

In the next section we perform the numerical calculation using DMRG
algorithm. The DMRG prefers OBC to PBC. Therefore we prepare the appropriate order
parameter for the discussion of the above models, which should show
whether the ground state of the system is dimerized.
Note that for the case of OBC, 
if we take the Hamitonian as,
\beqa
{\cal H}_{{\rm eff}}&=&\frac{J_{1}}{3}
\sum_{i=1}^{L-1}H_{{\rm eff}}^{i,i+1}(\alpha)+\frac{J_{1}}{6}
\sum_{i=1}^{L-2}H_{{\rm eff}}^{i,i+2}(\alpha) \nonumber\\
&=&\frac{J_{1}}{3}\sum_{i=1}^{L-2}(P_{i,i+1,i+2}-\frac{3}{8}(1+\alpha))+\frac{J_{1}}{6}
(H_{{\rm eff}}^{1,2}(\alpha)+H_{{\rm eff}}^{L-1,L}(\alpha)),\label{effMG2}
\enqa
the ground state is not degenerate for the finite-size system because 
of the last braketted terms.
$|\psi_{2}\rangle$
has higher energy than $|\psi_{1}\rangle$ with $\frac{1}{4}(1+\alpha)J_{1}$ for all finite system size.
Define the local Hamiltonian operator,
\beqa
H_{{\rm local}}^{i,i+1}=H_{{\rm eff}}^{i,i+1}(\alpha)-\frac{1}{L-1}
\sum^{L-1}_{j=1}\langle H_{{\rm eff}}^{j,j+1}(\alpha)\rangle.\label{local}
\enqa
In the three-leg ladder case, we replace $H_{{\rm eff}}^{i,i+1}$ with
$\sum_{n=1}^{3}
(H_{n,n}^{i,i+1}+H_{n,n+1}^{i,i}+H_{n,n+1}^{i+1,i+1})$. 
For the Hamiltonian (\ref{effMG2}), the expectation value of this operator
apparently has long-range order,  i.e.,
\beqa
\langle H^{i,i+1}_{{\rm local}}\rangle &=& (-1)^i \frac{3}{8}(1+\alpha),\nonumber
\enqa
where we assume $L\equiv 2$  mod  4.   
Therefore non-vanishing expectation value of this operator toward the
center of the system explicitly means that the system has dimerized ground state,
and breaks the translational symmetry in the thermodynamic limit.
On the other hand, in the case of the open ladder (effectively this model 
is considered as $S=1/2$ AFHC with OBC), the expectation value of this operator is
decaying as $(-1)^{r}/\sqrt{r}$ (up to logarithmic correction), where $r$ is distance from the
boundary.  This is because the scaling dimension of the dominant
part of the local Hamiltonian  is one-half.

\section{Numerical Calculations}
In this section we perform numerical calculations using DMRG algorithm
to show the original three-leg ladder system has the same nature
with the exactly solvable model, such as a non-zero excitation gap, a finite 
correlation length of the spin-spin correlation function and a doubled unit
cell of the ground state wave function. In the following calculations, we use 
the finite-size algorithm of DMRG which is necessary for the system with
a large number of degrees of freedom in a single-site block. (The
number of degrees of freedom for
the three-leg ladder is $2^{3}=8$, and that for the effective Hamiltonian
is $2^{2}=4$.)
We also show the result of  $S=1/2$ AFHC (effective
Hamiltonian of the open ladder) or the open
ladder for comparison. Note that $S=1/2$ AFHC and the open
ladders are expected to have power-low decaying correlation functions.

First we show the result of the effective Hamiltonian (\ref{hameff}).
In Fig. \ref{fig-1} we check the truncation-dependent accuracy by calculating 
triplet excitation gap ($S_{{\rm total}}^{z}=1,\tau_{{\rm
total}}^{z}=0$) in the case of truncation numbers $m=60$ and 80
for the finite-size algorithm.
By the figure we can see that
$m=80$ for the finite-size algorithm is enough for the calculation.
It looks also that there is an excitation gap. The inset is the data
for $m=80$ and the fitting function $0.277+0.276\exp (-0.0387 L)$. 
The extrapolated value of an excitation gap in the thermodynamic
limit is 0.27(7). 
In Fig. \ref{fig-2} we plot the spin-spin and
$\tau^{z}-\tau^{z}$ correlation functions for the system size
$74$ ($m=80$).
Both of them are exponentially decaying.
The correlation length is estimated as 2.7(4).  
In Fig. \ref{fig-3} we show the expectation value of the local Hamiltonian operator (\ref{local}).
The figure shows that the expectation value is not decaying.
This is the strong evidence to show that the ground state of the system is dimerized and breaks
the translational symmetry in the thermodynamic limit. To be convinced
more carefully that all the system with $0\leq J_{2}/J_1\leq 1/2$,
$J_{\perp}\rightarrow \infty$
is in the same phase, we see the low energy spectra of the effective
model with PBC by numerical exact 
diagonalization up to $12$ sites (Fig. \ref{fig-4} shows 12 sites spectra for
$J_2/J_1=0,0.25$ and 0.5). There is no level crossing
which supports no transition between $J_2/J_1=0$ and 0.5.
Secondly we show the result of the three-leg ladder. 
We pick up the typical value of $ J_{\perp}=1,5$ and  $10$ and the
system size is $22\times 3$.  
In Fig. \ref{fig-5}  spin-spin correlation functions are plotted.
They look like exponentially decaying. The correlation length tends
to increase as $J_{\perp}$ decreases. We also check the truncation error by
comparing the spin-spin correlation functions for truncation number
$m=40$, $60$ and $80$.(Inset of Fig. \ref{fig-5}) On the other hand, the data for 
the open ladders are power-law decaying as expected showing the system
is critical. 
We show the data for the expectation value of the local
Hamiltonian operator in Fig. \ref{fig-6}. These values do not decay, expressing
long-range dimer-order. These results suggest that the cylindrical ladder is essentially 
represented by the strong coupling model (Hamiltonian(\ref{hameff})) as expected by the
renormalization group analysis. On the other hand, for the open
ladders the expected results showing that the systems belong to the
universality class of level-1 $SU(2)$ WZW model are obtained. 
Finally in Fig. \ref{fig-7} we show the extrapolated values of the triplet excitation gap of the
cylindlical ladder for typical values of $J_{\perp}/J_1$
using data up to
$22\times 3$ sites.
The fitting function is deduced from the effective Hamiltonian
(Fig. \ref{fig-1}),
which is $a+b\exp(-cL)$, where $a,b,$ and $c$ are fitting constants. 

\section{Higher Number of Odd Legs Cylindrical Ladders}
In this section we comment on cylindrical ladders with a higher odd
number of legs in the strong coupling limit
($J_{\perp}\rightarrow\infty$). 
The ground states of an odd-sites antiferromagnetic ring are four-fold
degenerate.
(Though there is no exact proof of it, we expect this from 
the  exact numerical diagonalization of small system size and field
theory for a large number of system size. See ref. \citen{Eggert}.)
We can derive the strong coupling effective Hamiltonian of the $N$-leg
($N=$ odd) cylindrical ladder (See Appendix for derivation of it.) which 
is given as,
\beqa
{\cal H}_{{\rm eff}}^{N}=\frac{J_{1}}{N} \sum_{i}H_{{\rm eff}}^{i,i+1}(\alpha).\label{hameffn}
\enqa
Here $\alpha$ is positive and dependent on the number of legs $N$. For example 5-leg case
$\alpha=64/9$.  To calculate analytically the $N$-dependence
of $\alpha$ is too complicated for a higher number of legs.  For small
numbers of legs, however, we easily determine $\alpha$ numerically from 
the numerical exact diagonalization of system size $N\times 2$. In
Fig. \ref{fig-8} we plot the extrapolated value of $\alpha$ for
$N=3,5,7,9$ and $11$ using data up
to $J_{\perp}=2048$. The figure shows that $\alpha$ increases as the number 
of legs increases. In Fig. \ref{fig-9}, we plot the 
spin-spin
correlation function and the local Hamiltonian operator with $\alpha$
corresponding to $N=5$ and 7. The system size and truncation number are 
$L=50$ and $m=80$, respectively.  
These figures show that the
effective models for  cylindrical ladders of $N=5,7$ legs  have
dimerized ground states and  it suggests that there is an excitation gap.

\section{Summary and Discussion}
We study the ground state properties of the antiferromagnetic Heisenberg model on an odd number of
legs cylindrical ladder.
This model has frustrated interaction in the rung direction. The
frustration induces the dimerization and the model has an excitation
gap, that is different from the open ladder (which is with the open boundary condition in
the rung direction.). The essential point distinguishing the cylindrical ladder from 
the open ladder seems to be the degeneracy of the ground state of odd-site
antiferromagnetic Heisenberg chain 
 i.e. the number of freedom of single site in the strong
coupling effective Hamiltonian. 
First we consider the three-leg cylindrical ladder.
We reduce the effective Hamiltonian (\ref{hameff}) of the strong
coupling limit ($J_{\perp}\rightarrow \infty$), which is expected to be the fixed point of the
renomalization group treatment.  We introduce the next-nearest neighbor coupling 
($J_2$) for the effective Hamiltonian. For the model with special coupling ($J_1=2 J_2$), we can construct the exact 
ground state wave functions, which is direct products of two-site system wave
function. This construction is closely analogous to Majumdar-Ghosh
model. The wave function is apparently
dimerized. The exactly solvable model is expected to have an
excitation gap because the wave function is site-local.
To confirm that the original system ($J_2$=0) has same
nature with the exactly solvable model, we perform the numerical
calculations using DMRG algorithm for the effective model and the
three-leg cylindrical ladders. Adding to the spin-spin correlation function and triplet excitation gap, we calculate the expectation value of the local Hamiltonian
operator which expresses the ground state  being dimerized.
The result shows that the three-leg cylindrical ladders are
dimerized and  essentially expressed by the strong coupling effective model.

Lastly we briefly study the  cylindrical
ladders with a higher odd number of legs.
We can calculate the strong coupling effective Hamiltonian for the general number of legs,
which is the same form with the three-leg ladder. The $N$-dependence is
inserted into the parameter $\alpha$. We have already known the exact
solution for general $\alpha$ if we add the next-nearest neighbor
interaction $J_2 (=J_1/2)$. We expect the
dimerization  occurs for general $N$ even in the case of $J_2=0$. We also confirm this expectation
using DMRG calculations for small number of legs. 
We should comment
on the ladder with a large odd number of legs. From the Fig. \ref{fig-8} we 
see that positive parameter $\alpha$ is monotonous as a
function of $N$ up to $N=11$, we, therefore, expect that $\alpha$ for large number
of legs are between 4 and $\infty$. 
It is likely from  Fig. \ref{fig-9}
that at least in the strong coupling limit the finite
number of odd legs cylindrical ladder is dimerized in the ground state.
The calculation in \S 4  is the strong  coupling limiting case, therefore, in
this study  we cannot
mention about the connection between 2-dimensional isotoropic
Heisenberg model (which has Neel order and gapless spin wave
excitation) and the effective model studied here.

\section*{Acknowledgements}
The numerical work was partly based  on the program packages TITPACK
ver.2 by H.Nishimori.
The computation in this work were done partially  on the FACOM VPP500 in the Supercomputer Center of ISSP, University
of Tokyo. 
This research is supported 
in part by Grants-in-Aid for Scientific Research Fund from the Ministry of 
Education, Science and Culture (08640445).
 K.K. are supported by JSPS Reserch Fellowships for Young Scientists.

\appendix
\section{The Effective Hamiltonian for General $N$}
In this Appendix, we derive the effective Hamiltonian (\ref{hameffn}) for general $N$.
As denoted in \S 4, the ground states of odd-sites
antiferromagnetic ring are four-fold degenerate (two doublets) and these two doublets 
have opposite sign of momentum i.e. these momenta are $k$ and
$-k$.
We write these four states as  $|\uparrow L \rangle$, $|\downarrow L
\rangle$, $|\uparrow R \rangle$ and $|\downarrow R \rangle$.
Here $L$ and $R$ represent the momentum $k$ and
$-k$ ($k\neq 0$), and we set $\omega=\exp ({\rm i} k)$. Note that interchanging $L$
and $R$
corresponds to interchanging  $\omega$ and $\omega^{-1}$.
Matrix elements of $S_{n,i}^{z}$ and $S_{n,i}^{+}$ in the restricted
Hilbert space are 
\beqa
\langle\uparrow L|S_{n,i}^{z}|\uparrow L \rangle=a_{n},&&\quad \langle\uparrow
R|S_{n,i}^{z}|\uparrow R \rangle=a_{n}^{*}, \nonumber\\
\langle\uparrow L|S_{n,i}^{z}|\uparrow R \rangle=b_{n},&&\quad \langle\uparrow
R|S_{n,i}^{z}|\uparrow L \rangle=b_{n}^{*}, \nonumber\\
\langle\uparrow L|S_{n,i}^{+}|\downarrow L \rangle=c_{n},&&\quad \langle\uparrow
R|S_{n,i}^{+}|\downarrow R \rangle=c_{n}^{*}, \nonumber\\
\langle\uparrow L|S_{n,i}^{+}|\downarrow R \rangle=d_{n},&&\quad \langle\uparrow
R|S_{n,i}^{+}|\downarrow L \rangle=d_{n}^{*}, \nonumber
\enqa
where $n$ is the site number of odd site ring and $a^*$ represents the 
complex conjugate of $a$. It is obvious that
interchanging $\uparrow$ and $\downarrow$ is interchanging the sign of 
$a_{n}$ and $b_{n}$ and setting $c_{n}=d_{n}=0$. We set $T$ is one-site shift operator for the $n$-direction, i.e. $T^{m}|\uparrow L
\rangle=\omega^{m}|\uparrow L \rangle$ and $T^{m}|\uparrow R
\rangle=\omega^{-m}|\uparrow R \rangle$. Then we conclude using the relation $T^{m}S^{z}_{n,i}T^{-m}=S^{z}_{n+m,i}$ that 
\beqa
a_{n}=a_{1}, \quad b_{n}=b_{1} \omega^{-2(n-1)}, \quad  c_{n}=c_{1},
\quad d_{n}=d_{1} \omega^{-2(n-1)}.\label{abcd}
\enqa
The operator  $S^{z}_{i}$ of the effective Hamiltonian is the sum of $S^{z}_{n,i}$. Then using
the relation eqs.(\ref{abcd})  we know $a=1/2N$.
Setting the spin reversal operator $P$ as
$\prod_{n=1}^{N}2S^{x}_{n,i}$, we can derive the relations,
\beqa
c_{1}=\langle\uparrow L|S_{1,i}^{+}|\downarrow L \rangle=\langle\downarrow
L|P S_{1,i}^{+} P|\uparrow L \rangle=\langle\downarrow L|S_{1,i}^{-}|\uparrow L \rangle=c_{1}^{*},
\enqa
i.e.  $c_{1}$ is also an real number.
We also note that $\sum_{n=1}^{N} b_{n}^{2}=\sum_{n=1}^{N} b_{n}=\sum_{n=1}^{N}
d_{n}^{2}=\sum_{n=1}^{N} d_{n}=0$. Finally we obtain the effective Hamiltonian of
the cylindrical ladder for general $N$,
\beqa
{\cal H}_{{\rm eff}}^{N}=&&J_{1} \sum_{i}\left\{ 4 S_{i}^{z}S_{i+1}^{z}\sum_{n=1}^{N}\left( a_{n}^{2}+|b_{n}|^{2}(\tau ^{+}_{i}\tau
^{-}_{i+1}+ \tau ^{-}_{i}\tau
^{+}_{i+1})\right) \right.\nonumber\\
&&\left.\frac{1}{2}(S_{i}^{+}S_{i+1}^{-}+S_{i}^{-}S_{i+1}^{+})\sum_{n=1}^{N}\left( c_{n}^{2}+|d_{n}|^{2}(\tau ^{+}_{i}\tau
^{-}_{i+1}+ \tau ^{-}_{i}\tau
^{+}_{i+1})\right) \right \}.\label{hampri}
\enqa
In the derivation above, the $SU(2)$ symmetry of the each interaction  $\mbox{\boldmath $S$}_{n,i} \cdot
\mbox{\boldmath $S$}_{n,i+1}$ is preserved, so it should be that the
effective Hamiltonian (\ref{hampri}) also possesses the symmetry. This
means that $4 a^2_1=c^2_1$ and $|b_{1}|^2/a^2_1=|d_{1}|^2/c^2_1$. Then we
reached the result (\ref{hameffn}) with $\alpha=|b_{1}|^2/a^2_1$ ($>0$).

We calculate $\alpha$ for $N=5$ as follows.
The ground states for 5-sites ring ($H=\sum_{n=1}^{5}\mbox{\boldmath $S$}_{n,i}\cdot
\mbox{\boldmath $S$}_{n+1,i} $) with energy $-\frac{1}{4}(2\sqrt{5}+3)$ are 
\beqa
|\uparrow L
\rangle=&&\frac{1}{\sqrt{|\beta_{1}|^2+|\beta_{2}|^2}}(\beta_{1}|{\rm 
 I}_{1}\rangle+\beta_{2}|{\rm II}_{1}\rangle),\nonumber 
\enqa
where $\beta_{1}=(\omega^{-1}+\omega^{-2})/2$,
$\beta_{2}=-(\omega^{-1}+\omega)-\frac{3}{2}$
and $\omega=\exp(\frac{2}{5}{\rm i}\pi)$. The vectors $|{\rm I}_{1}\rangle $
and $|{\rm II}_{1}\rangle $ are
\beqa
|{\rm I}_{1}\rangle&=&\frac{1}{\sqrt{5}}\left
( |\uparrow\uparrow\uparrow\downarrow\downarrow\rangle
+\omega|\uparrow\uparrow\downarrow\downarrow\uparrow\rangle
+\omega^2|\uparrow\downarrow\downarrow\uparrow\uparrow\rangle
+\omega^{-2}|\downarrow\downarrow\uparrow\uparrow\uparrow\rangle
+\omega^{-1}|\downarrow\uparrow\uparrow\uparrow\downarrow\rangle \nonumber\right) \nonumber\\
|{\rm II}_{1}\rangle&=&\frac{1}{\sqrt{5}}\left
( |\uparrow\downarrow\uparrow\downarrow\uparrow\rangle
+\omega|\downarrow\uparrow\downarrow\uparrow\uparrow\rangle
+\omega^2|\uparrow\downarrow\uparrow\uparrow\downarrow\rangle
+\omega^{-2}|\downarrow\uparrow\uparrow\downarrow\uparrow\rangle
+\omega^{-1}|\uparrow\uparrow\downarrow\uparrow\downarrow\rangle \nonumber\right).
\enqa 
Other three states $|\downarrow L\rangle$, $|\uparrow R\rangle$ and
$|\downarrow R\rangle$ are produced by interchanging $\uparrow$ and
$\downarrow$ or taking complex conjugate.
After calculations we get,
\beqa
b_{1}=\frac{-2\omega^{-1}+3\omega-3\omega^2}{5(|\beta_{1}|^2+|\beta_{2}|^2)}.
\enqa
This leads to the result $\alpha=64/9$.

\pagebreak

\pagebreak
%\Figures
{\bf Figure captions}

\begin{figure}

\caption{The finite size triplet excitation gap for the effective
  Hamiltonian ($L=74$). The data is for the truncation number $m=
  60$($\bigcirc$) and 80($\times$). The inset is the data
for $m=80$ and the fitting function $0.277+0.276\exp (-0.0387 L)$.}
\label{fig-1}
\end{figure}

\begin{figure}

\caption{The correlation functions $(-1)^r \langle
  S^{z}_{L/2}S^{z}_{L/2+r}\rangle$ ($\Box$) and
  $(-1)^r\langle\tau^{z}_{L/2}\tau^{z}_{L/2+r}\rangle$ ($+$)
for the effective
  Hamiltonian with system size $L=74$ and truncation number $m=80$ are
  plotted in semi-log scale. The lines are for the  guide to eyes
  ($1.58\times 10^{-2}\exp(-r/2.74)$ and $3.74\times 10^{-3}\exp(-r/2.74)$).}
\label{fig-2}
\end{figure}
\begin{figure}

\caption{The expectation value of the local Hamiltonian operator 
$(-1)^r\langle H^{r,r+1}_{{\rm local}}\rangle$ (The symbol +, $1\leq r\leq L/2$) is
plotted in log-log scale for the effective Hamiltonian of
the same $L$ and $m$ with Fig.2. 
The symbol $\Diamond$ is for $S=1/2$ AFHC and the line is 
$0.20/\protect{\sqrt{r}}$.}

\label{fig-3}
\end{figure}

\begin{figure}

\caption{The low-energy spectra for the effective Hamiltonian in the
  periodic boundary condition with $J_{2}/J_{1}=0,0.25$ and $0.5$. The
  system size $L$ is 12. x-axis is lattice momentum in unit
  $2\pi/L$. The symbols $\Box$, $\times$, are spin-singlet with
  $\tau^{z}_{{\rm total}}=0$ and $\tau^{z}_{{\rm total}}=1$ (The
  blacked symbol represents lowest two spin-singlet with
  $\tau^{z}_{{\rm total}}=0$.), the
  symbols $\triangle$, $\bigcirc$ are spin-triplet with $\tau^{z}_{{\rm total}}=0$ and $\tau^{z}_{{\rm total}}=1$, respectively.}
\label{fig-4}
\end{figure}
\begin{figure}

\caption{The correlation function $(-1)^r\langle
  S^{z}_{L/2}S^{z}_{L/2+r}\rangle$  for the three-leg ladders. (We use 
  semi-log scale for cylindrical ladders and log-log scale for open ladders.) The
  system size is $22 \times 3$ and truncation number $m=60$. We choose
  typical $J_{\perp}$ as 1 ($\Box$), 5 ($\triangle$) and 10
  ($\times$). The inset is the data for the cylindrical ladder with
  $J_{\perp}=5$ with $m=40(\times),60(\triangle),$ and $80$($\Diamond$), we can see $m=60$ is enough for the calculation.}
\label{fig-5}
\end{figure}
\begin{figure}

\caption{The expectation value of local Hamiltonian operator $(-1)^r\langle
  H^{r,r+1}_{{\rm local}}\rangle$ ($1 \leq r \leq L/2$) for three-leg
  open ladders (black) and cylindrical ladders (white) with $L=22,N=3$. (The symbols
  $\Diamond$, $\Box$ and $\bigtriangledown$ are for $J_{\perp}=1,5$ and 
  10, respectively. The line $0.20/\protect{\sqrt{r}}$ are guide to eyes.)}
\label{fig-6}
\end{figure}

\begin{figure}

\caption{The extrapolated value of the triplet excitation gap for the
  three-leg cylindrical ladders, using the data up to $22\times3$ with 
  $m=60$. The dotted line (0.27(7)) is the gap of the
  effective Hamiltonian($J_{\perp}\rightarrow\infty$).
}
\label{fig-7}
\end{figure}
\begin{figure}
\caption{The extrapolated value of $\alpha$ for the higher number of legs cylindrical
  ladder using  the data of the numerical exact diagonalization of $N\times 2$ sites.}
\label{fig-8}
\end{figure}

\begin{figure}
\caption{The correlation function $(-1)^r\langle
  S^{z}_{L/2}S^{z}_{L/2+r}\rangle$ and the local  Hamiltonian
  operator $(-1)^r\langle H^{r,r+1}_{{\rm local}}\rangle$ ($1\leq r\leq L/2$) of
the effective Hamiltonian with  $N=3$ ($\Diamond$),5 ($\Box$) and 7
($\triangle$)
are plotted in semi-log scale.}
\label{fig-9}
\end{figure}

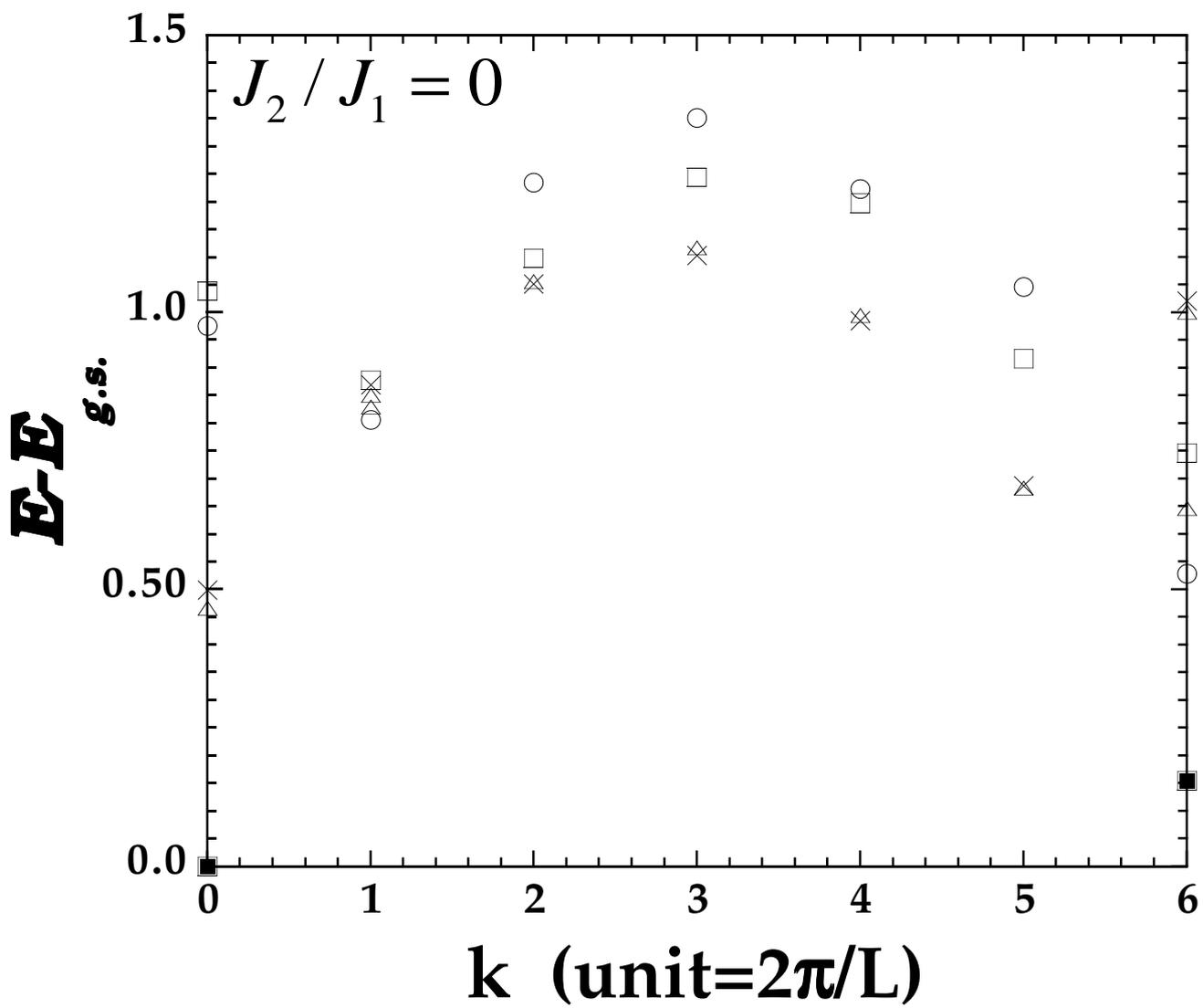
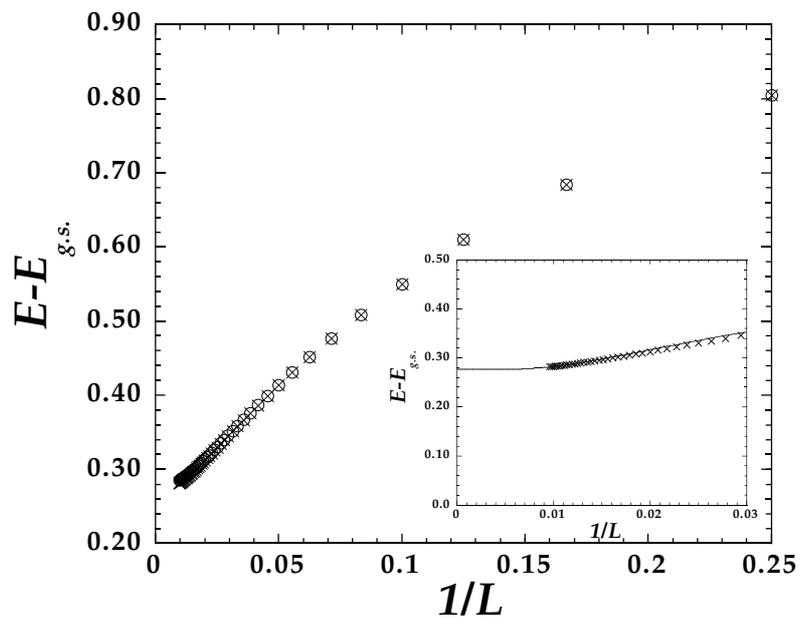
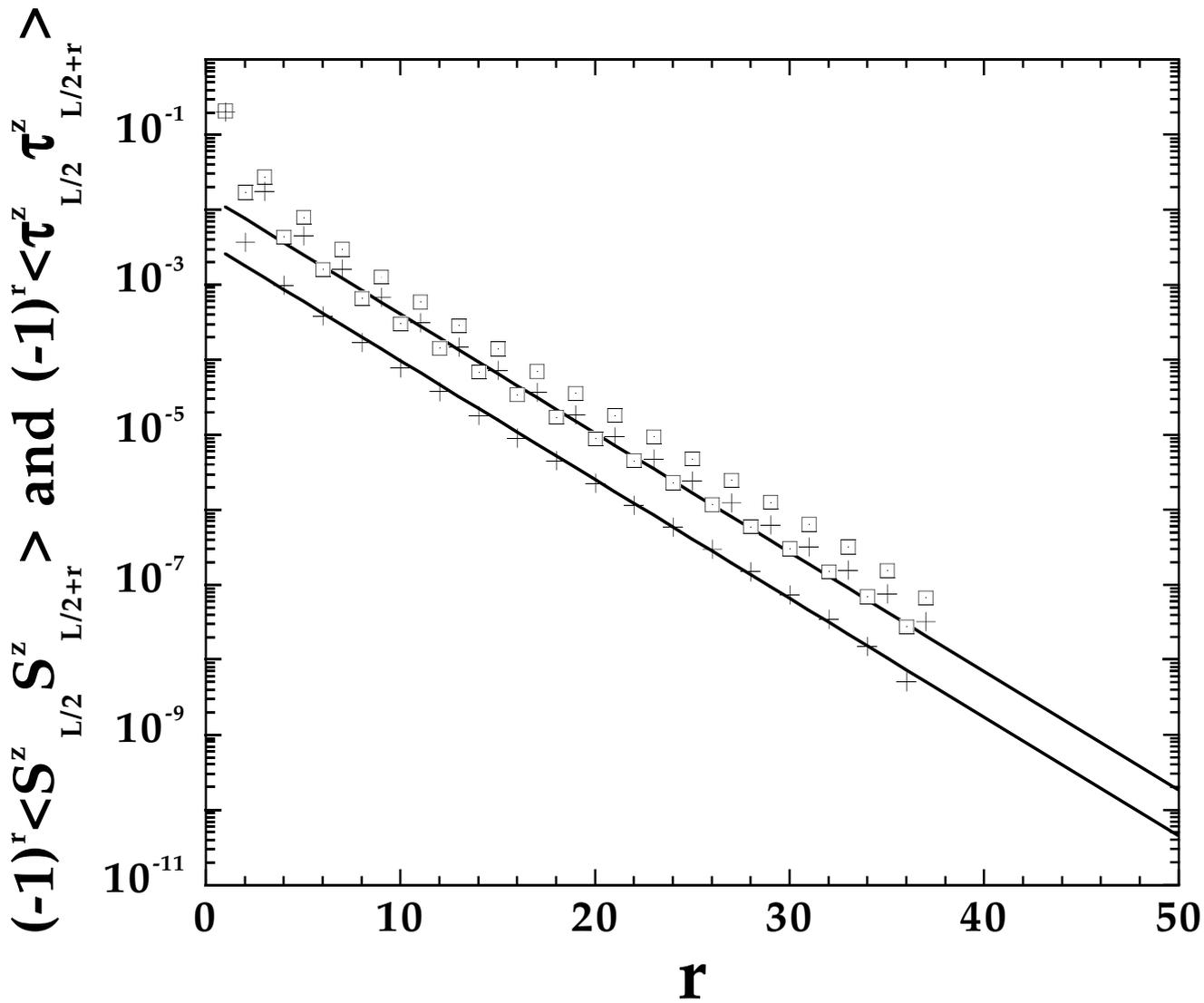
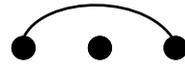
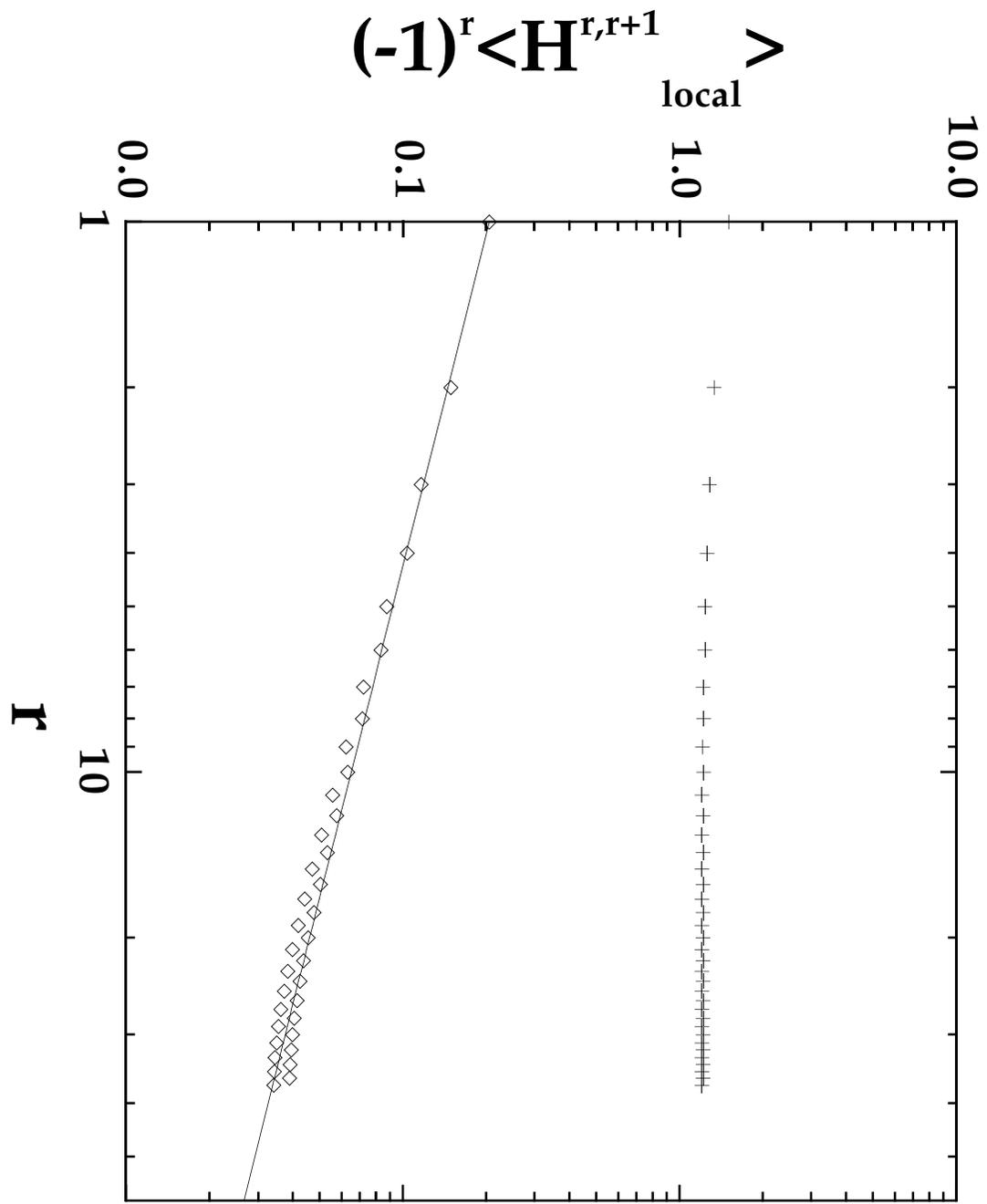
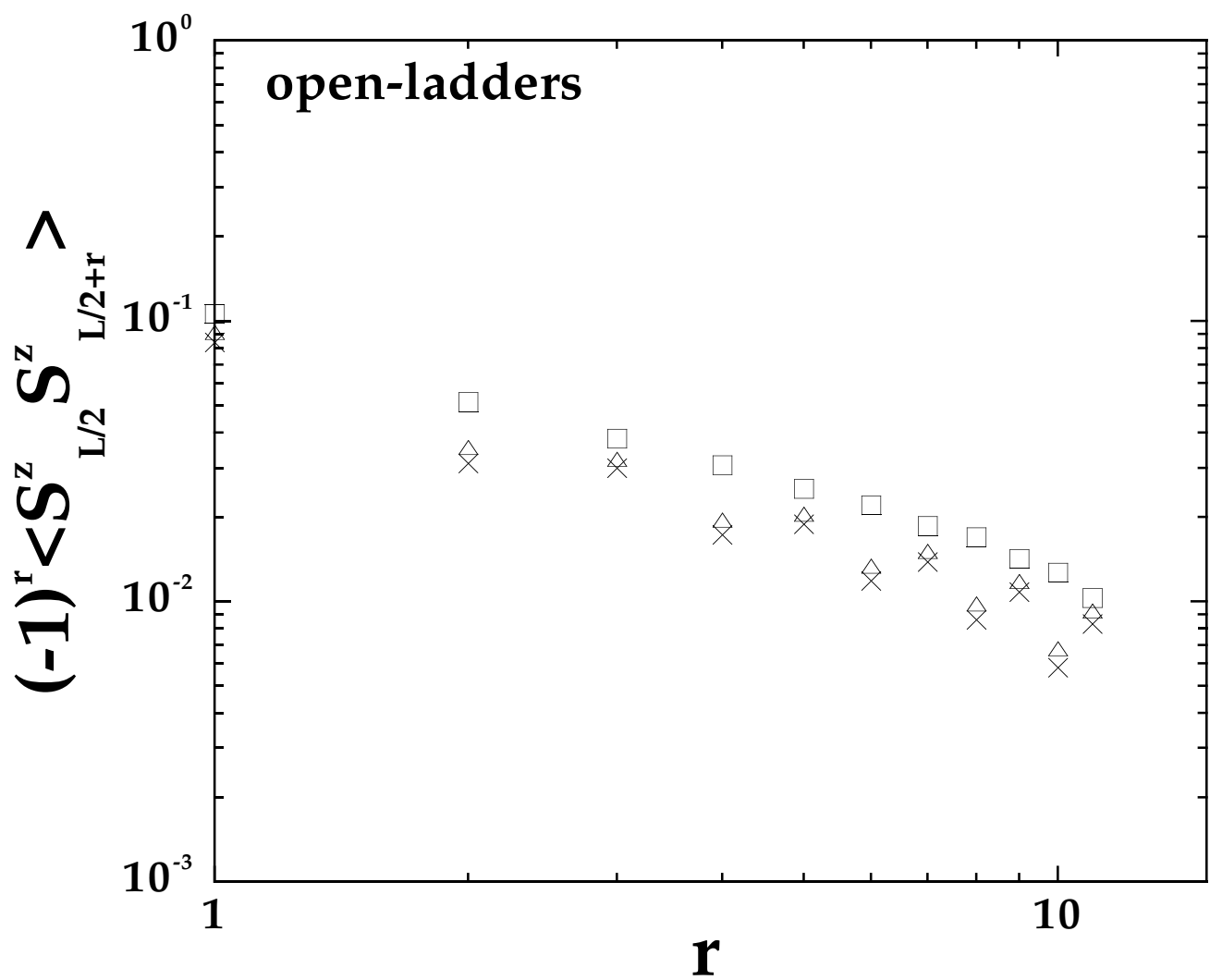
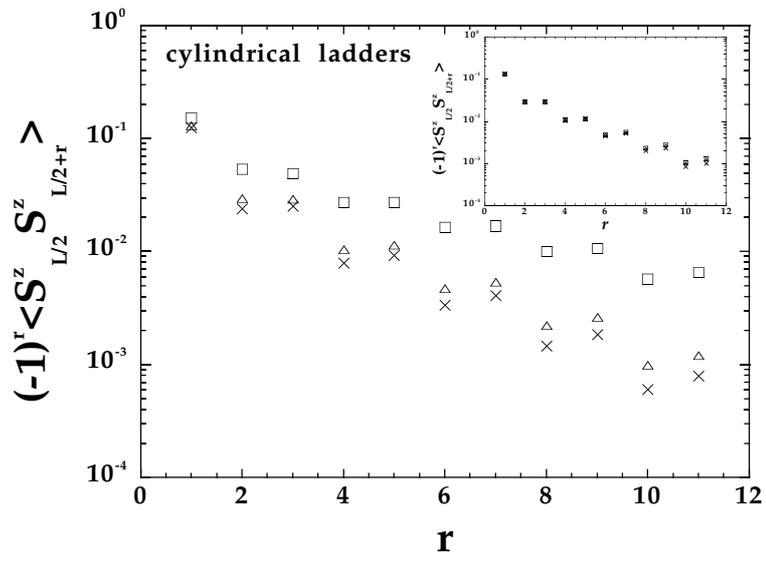
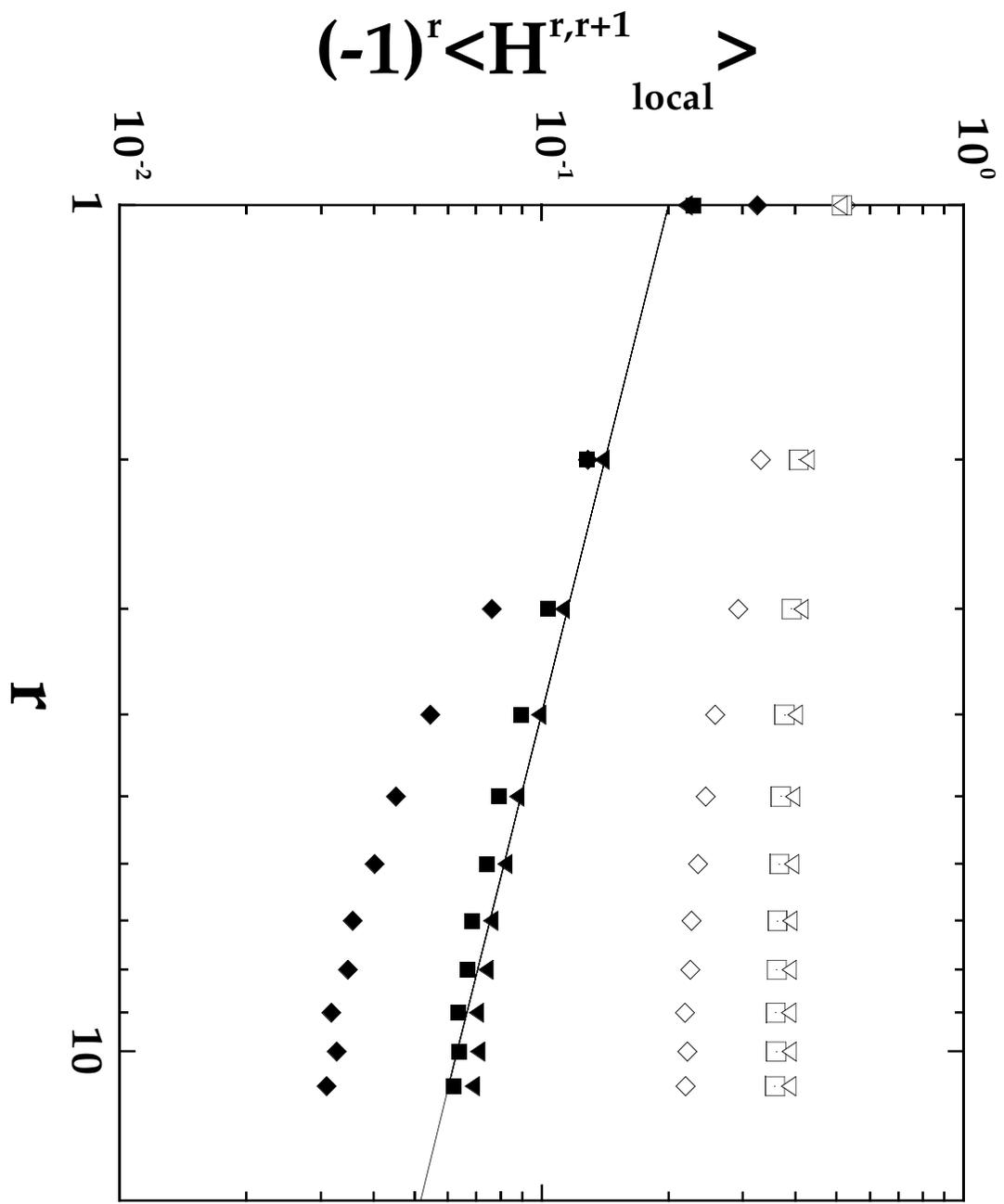
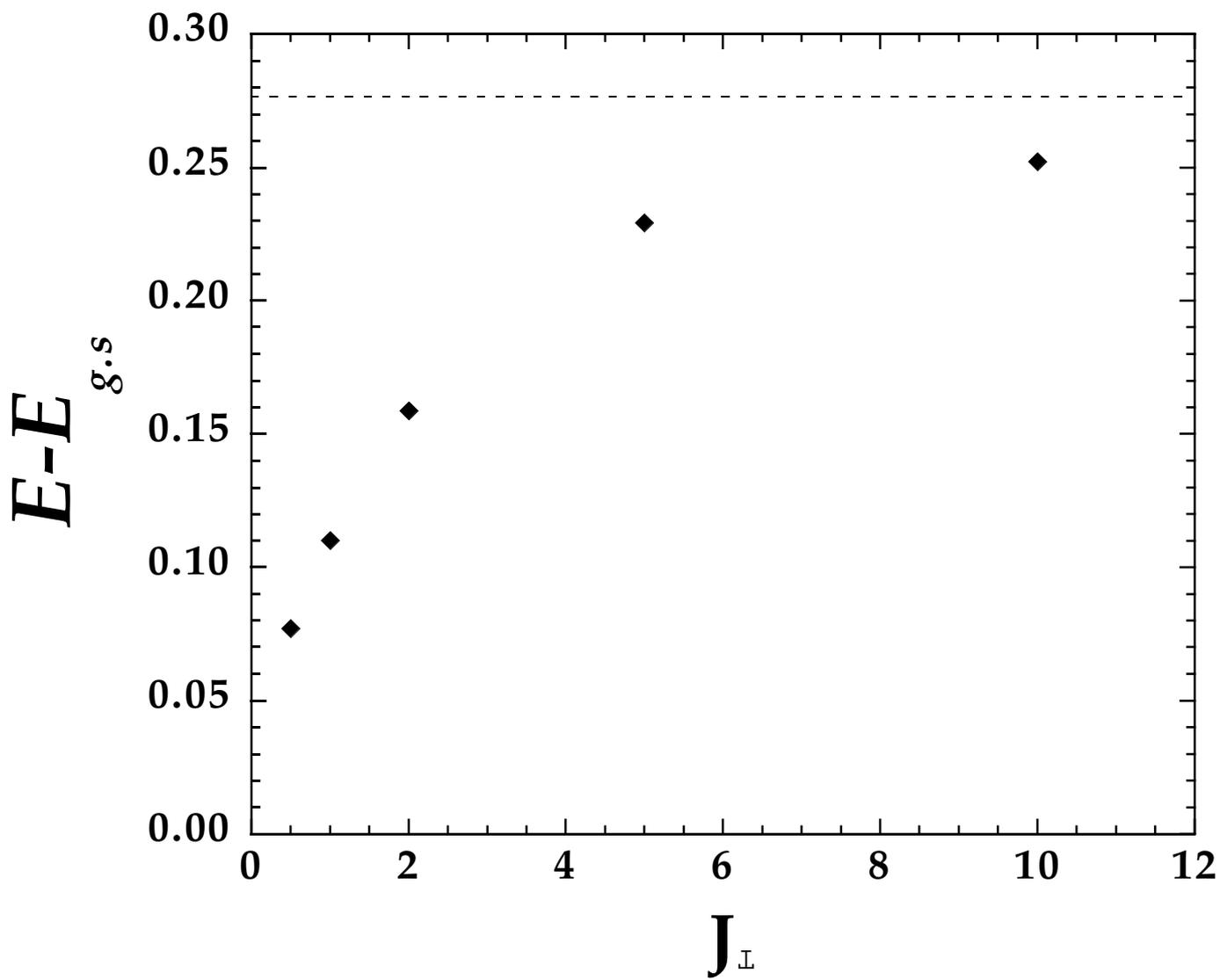
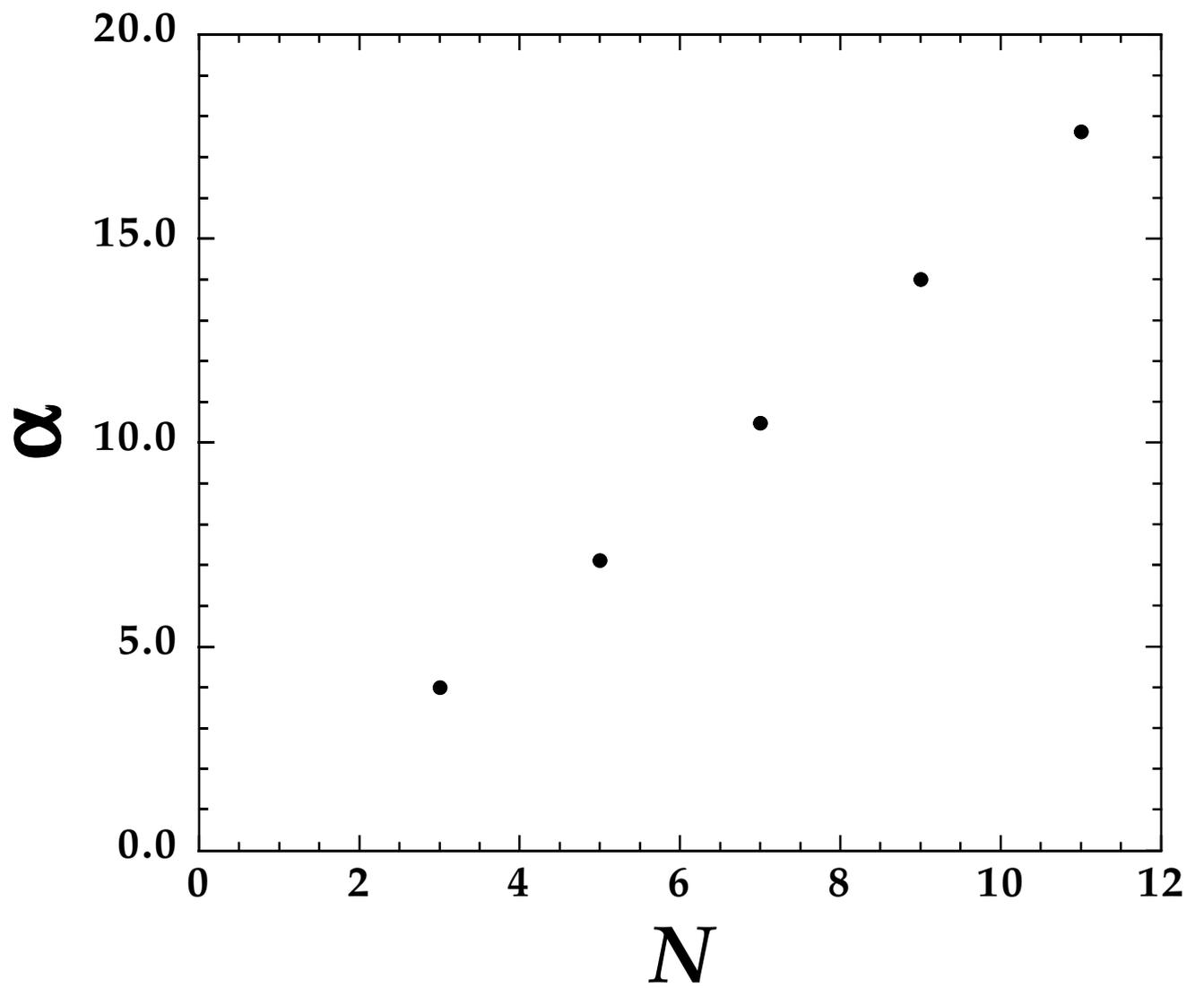
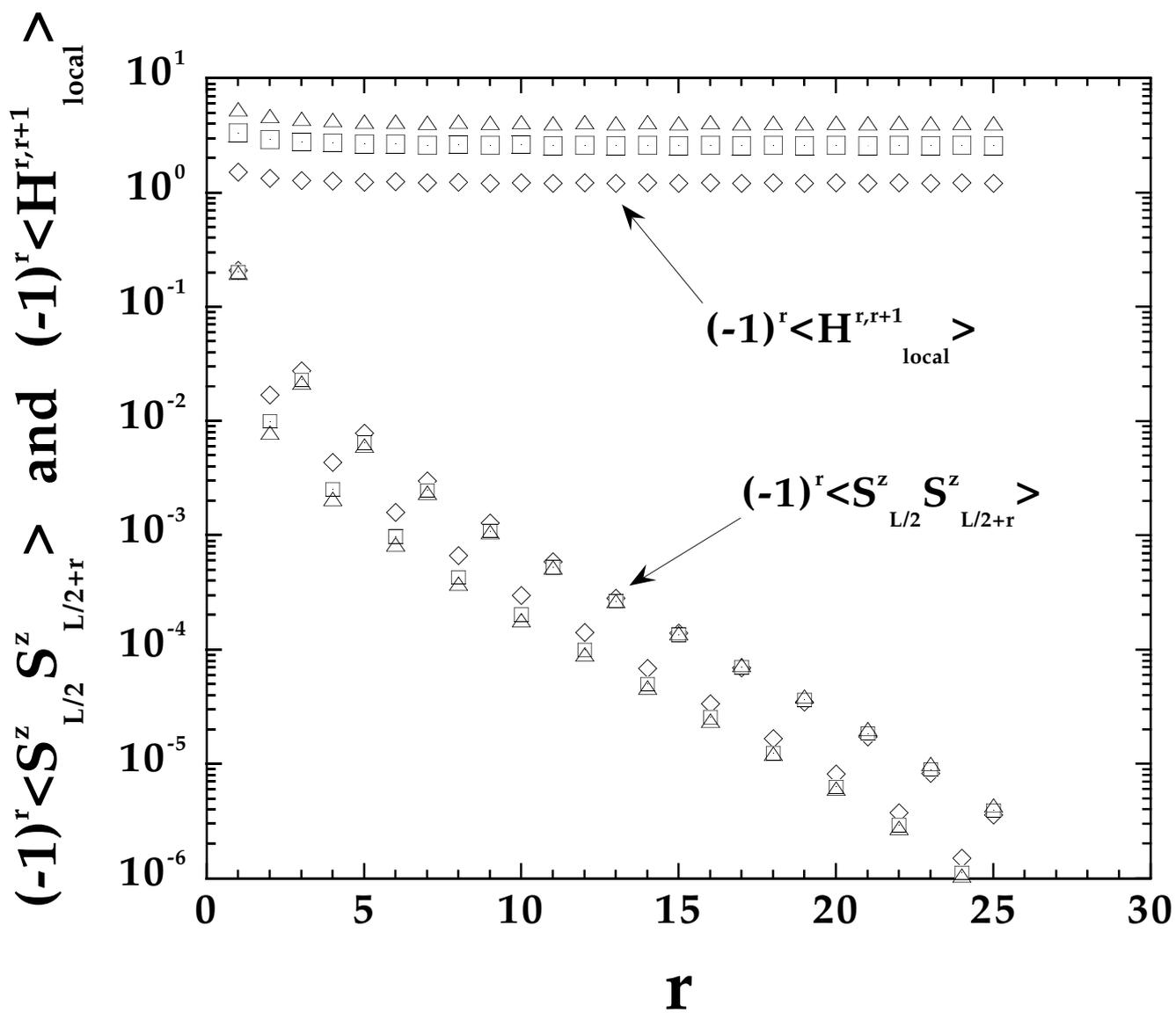
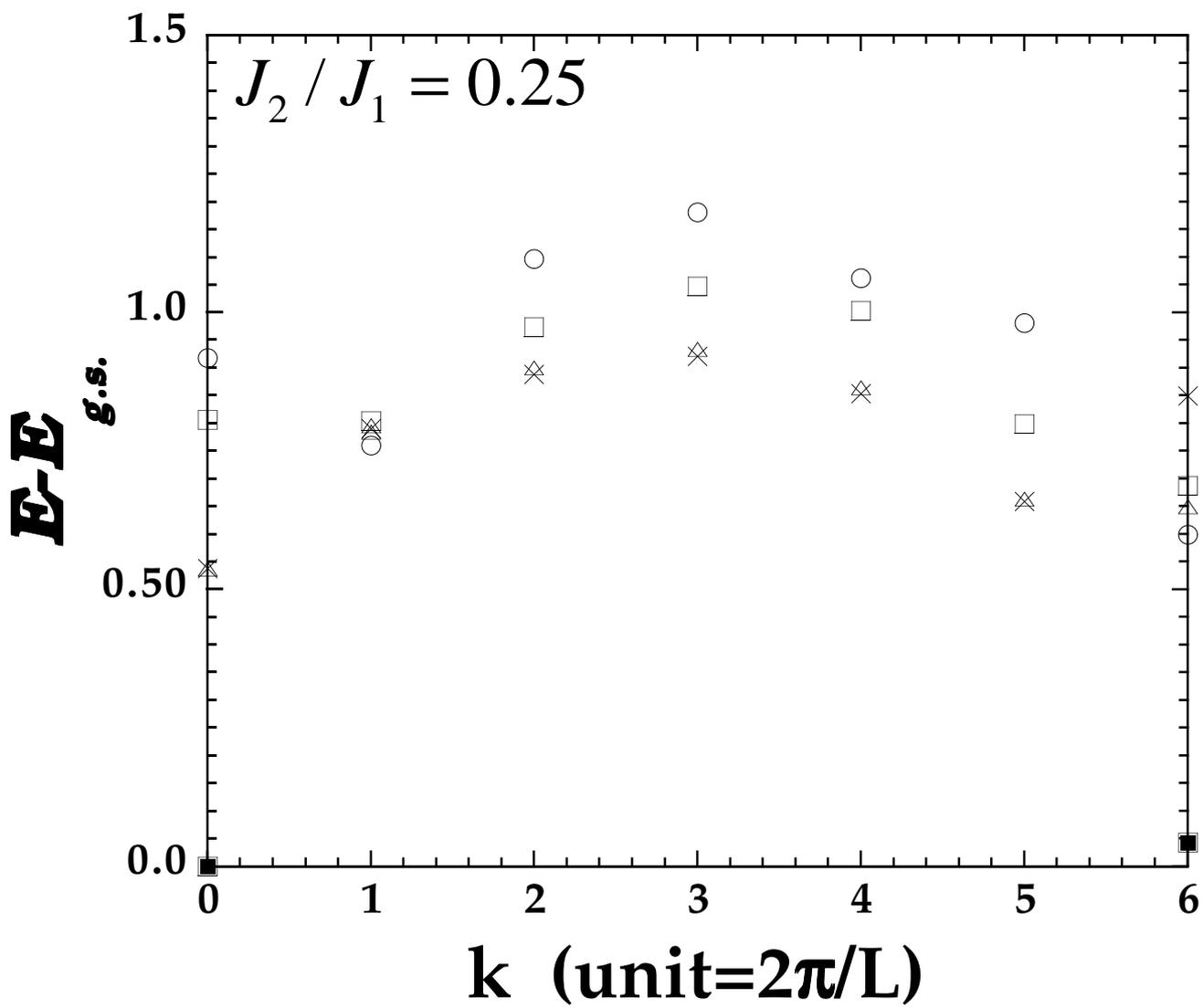
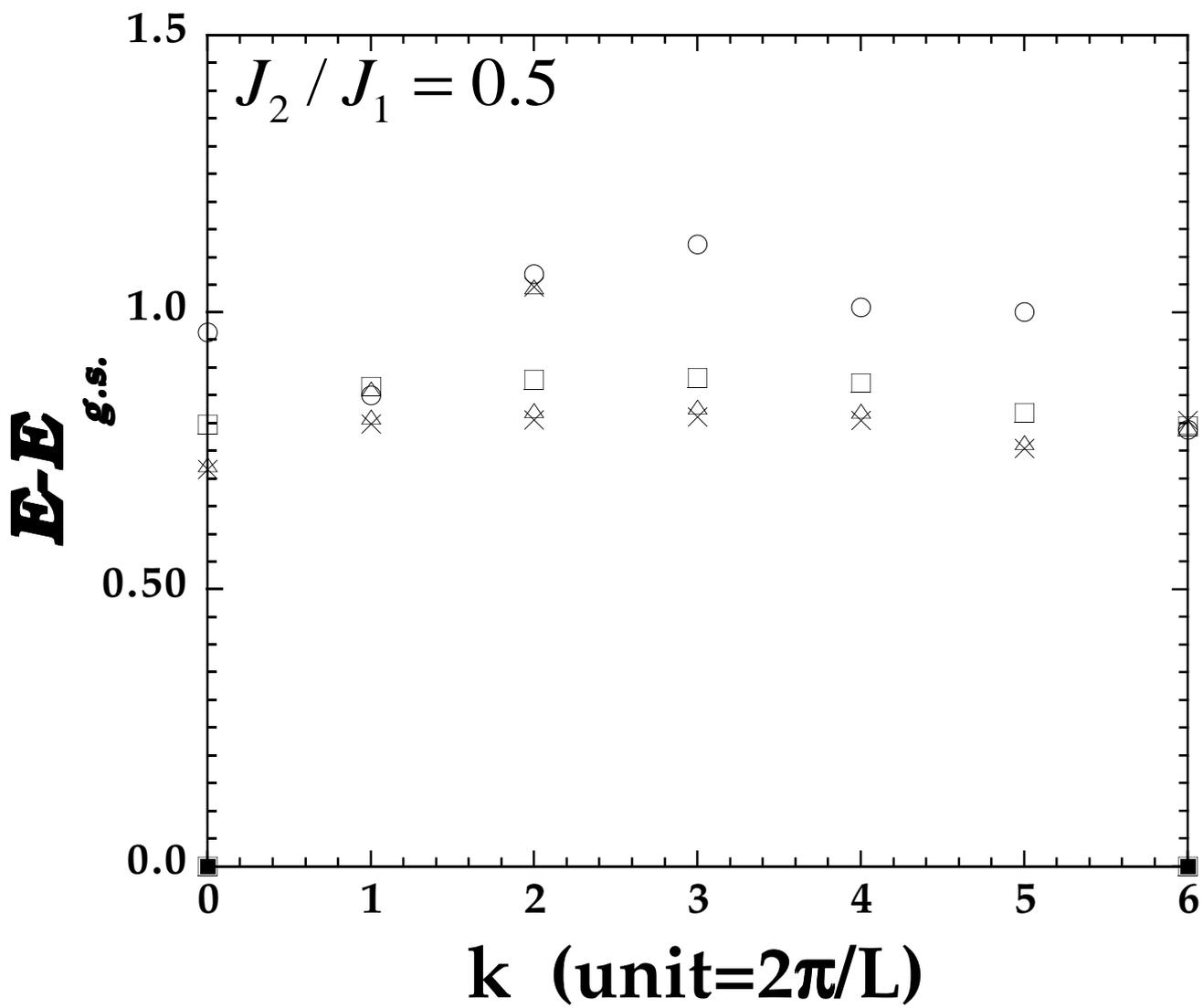
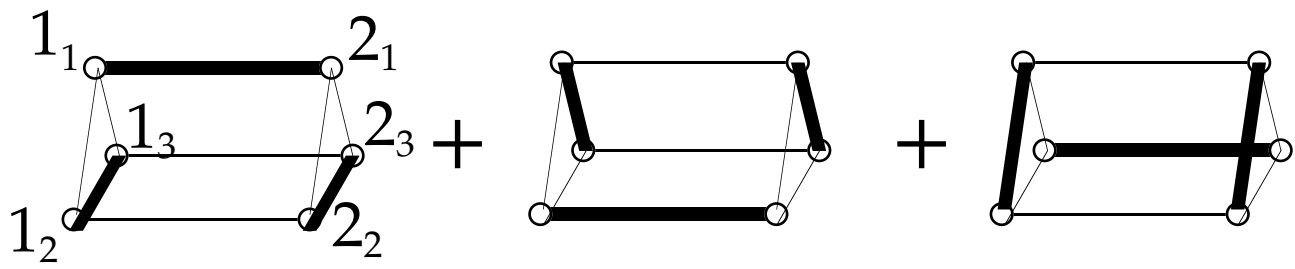
\end{document}